%% file: main.tex
\documentclass[conference]{IEEEtran}
\IEEEoverridecommandlockouts
\usepackage{amsmath,amssymb,amsfonts}
\usepackage{algorithmic}
\usepackage{graphicx}
\usepackage{textcomp}
\usepackage{xcolor}
\def\BibTeX{{\rm B\kern-.05em{\sc i\kern-.025em b}\kern-.08em
    T\kern-.1667em\lower.7ex\hbox{E}\kern-.125emX}}

\usepackage[font=small]{caption}
\usepackage[caption=false,font=footnotesize]{subfig}
\usepackage{xspace}
\usepackage{microtype}
\usepackage{bm}

\usepackage{float}

\usepackage{multicol}
\usepackage{tabularx}
\usepackage{tabularray}
\usepackage{booktabs}
\usepackage{multirow}
\newcolumntype{L}[1]{>{\raggedright\arraybackslash}p{#1}}

\usepackage[numbers,sort&compress]{natbib}

\usepackage[hidelinks]{hyperref}
\newcommand{\appref}[1]{\hyperref[#1]{Appendix~\ref{#1}}}

\definecolor{cbone}  {HTML}{006BA4} %
\definecolor{cbtwo}  {HTML}{FF800E} %
\definecolor{cbthree}{HTML}{ABABAB} %
\definecolor{cbfour} {HTML}{595959} %
\definecolor{cbfive} {HTML}{5F9ED1} %
\definecolor{cbsix}  {HTML}{C85200} %
\definecolor{cbseven}{HTML}{898989} %
\definecolor{cbeight}{HTML}{A2C8EC} %
\definecolor{cbnine} {HTML}{FFBC79} %
\definecolor{cbten}  {HTML}{CFCFCF} %

\colorlet{primarycolor}{cbtwo}
\colorlet{primaryshaded}{cbsix}
\colorlet{secondarycolor}{cbone}
\colorlet{secondaryshaded}{cbeight}
\colorlet{tertiarycolor}{cbfour}
\colorlet{tertiaryshaded}{cbten}

\input{./commands.tex}
\renewcommand{\paragraph}[1]{{\vskip 6pt \noindent\textbf{#1}. }}

\input{./tikzpgf.tex}

\usepackage{pifont} %

\usepackage{fancyhdr}
    
\begin{document}

\title{The Impact of Uniform Inputs on Activation Sparsity and Energy-Latency Attacks in Computer Vision}

\author{\IEEEauthorblockN{Andreas Müller}
\IEEEauthorblockA{\textit{Ruhr University Bochum} \\
Bochum, Germany}
\and
\IEEEauthorblockN{Erwin Quiring}
\IEEEauthorblockA{\textit{Ruhr University Bochum} \\
Bochum, Germany}
\IEEEauthorblockA{\textit{International Computer Science Institute (ICSI)} \\
Berkeley, USA}
}

\maketitle

\begin{abstract}
Resource efficiency plays an important role for machine learning nowadays. The energy and decision latency are two critical aspects to ensure a sustainable and practical application. 
Unfortunately, the energy consumption and decision latency are not robust against adversaries. 
Researchers have recently demonstrated that attackers can compute and submit so-called sponge examples at inference time to increase the energy consumption and decision latency of neural networks.
In computer vision, the proposed strategy crafts inputs with less activation sparsity which could otherwise be used to accelerate the computation.   

In this paper, we analyze the mechanism how these energy-latency attacks reduce activation sparsity.
In particular, we find that input uniformity is a key enabler. %
A uniform image, that is, an image with mostly flat, uniformly colored surfaces, triggers more activations due to a specific interplay of convolution, batch normalization, and ReLU activation. 
Based on these insights, we propose two new simple, yet effective strategies for crafting sponge examples: sampling images from a probability distribution and identifying dense, yet inconspicuous inputs in natural datasets.
We empirically examine our findings in a comprehensive evaluation with multiple image classification models and show that our attack achieves the same sparsity effect as prior sponge-example methods, but at a fraction of computation effort.
We also show that our sponge examples transfer between different neural networks.
Finally, we discuss applications of our findings for the good by improving efficiency by increasing sparsity.

\end{abstract}

\fancypagestyle{firststyle}
{
	\fancyhf{}
	\chead{\small\textit{ ---------------  Accepted at the Deep Learning Security and Privacy Workshop (DLSP) 2024, co-located with IEEE S\&P --------------- }}	
	\renewcommand{\headrulewidth}{0.0pt} }
\thispagestyle{firststyle}

\section{Introduction}\label{sec:introduction}

Resource efficiency in deep learning is not a choice but a necessity. 
Recent breakthroughs such as Transformer-based models~\cite{VasShaPar17, BroManRyd+20} 
have led to impressive progress in fields such as natural language processing and computer vision, but at the cost of significant computational resources~\cite{PatGonLe21}, resulting in high energy consumption and decision latency. This leads to a considerable environmental and financial overhead, and also limits the deployment on edge devices such as smartphones.

As a result, industry and academia have extensively developed various methods to reduce the energy, latency, 
and memory requirements in machine learning~\cite{NikMahMos18, HanLiuMao16}. 
Leveraging sparsity is, in fact, one of the standard methods for achieving resource efficiency~\cite{KurKopGel20}. 
At inference time for example, it is possible to exploit 
\emph{dynamic activation sparsity} to decrease the energy and latency 
overhead by reducing memory requirements and ineffectual 
computations~\cite{JiaHuaYu23, JanLeeKim21, OhSoKim21, KurKopGel20, 
MahEdoZad20, ShaParSud18, ParRhuMuk17, Geo19}.

However, energy consumption and decision latency are not robust 
against malicious actors. Recently, \mbox{\citet{ShuZhaBat21}} has shown 
that attackers can generate \emph{sponge examples} to decrease the activation sparsity. This attack reduces the effectiveness of sparsity-aware acceleration methods, leading to an increase in energy consumption and latency per query.
Such an attack can be used for denial-of-service 
attacks, to slowdown time-critical applications such as autonomous 
driving~\cite{ShaZolDem23}, or to quickly drain battery-powered 
devices~\cite{MarHsiHa04} such as mobile devices powered by 
sparsity-aware NPUs~\cite{JanLeeKim21}. 

In this paper, we revisit energy-latency attacks that are targeting the activation sparsity. First, we analyze how sponge examples can reduce activation sparsity. 
We find that previous attacks generate images with uniformly-colored surfaces. We theoretically analyze that such uniform inputs are beneficial for an attack, because they cause narrow convolutional outputs around zero. These values are more likely to be moved to the positive ReLU domain due to batch normalization. This property is common across image classification models and explains the transferability of samples across models in terms of activation sparsity.
Based on these insights, we then derive two novel strategies to craft sponge examples. 
We empirically validate our analysis and evaluate our proposed attacks on various common vision models. Our attacks achieve the same density effect as prior work, but at a fraction of time.

Finally, our work discusses applications of our gained insights to improve sparsity in non-adversarial settings. If a dataset is rather uniform, we show that fine-tuning on such data distributions can improve activation sparsity.   

\paragraph{Contributions} In summary, we make the following contributions in this paper:
\begin{enumerate}
\setlength{\itemsep}{4pt}

\item {\em Analysis of Uniformity on Activation Sparsity.}
We examine the impact of input distribution on activation sparsity. We theoretically analyze why uniform inputs reduce sparsity and are thus beneficial for energy-latency~attacks. 

\item {\em Novel Attack Strategies.} 
We propose two simple strategies to craft sponge examples. The strategies achieve the same density effect as prior work, but at a fraction of time.   

\item {\em Comprehensive Evaluation.}
We evaluate the attack strategies on various classification models. We also show the transferability of sponge examples across models. 
\end{enumerate}

Our dataset and implementation are publicly available at \textcolor{secondarycolor}{\url{https://github.com/and-mill/2024-sponge-example-analysis}}.

\section{Background}\label{sec:background}
In this section, we briefly review normalization \& optimization in deep neural networks and then present the background on energy-latency attacks before bringing all concepts together in \autoref{sec:main}. 

\subsection{Notation}
For simplicity, we assume a single input and skip the batch dimension in the following. 
For a convolutional neural network, we denote the input at the layer~$l$ by $\Z^l \in \mathbb{R}^{C_l \times H_l \times W_l}$ with $C_l$ channels, height $H_l$, and width $W_l$. We refer to a specific channel~\ch by $\Z^l_c \in \mathbb{R}^{H_l \times W_l}$. 

\subsection{Normalization in Neural Networks}
Normalization is a standard component in machine learning. We shortly recap batch normalization which is prevalent in vision models and a key enabler for energy-latency attacks.

\paragraph{Batch Normalization}\label{sec:batchnorm}
For intermediate layers, batch normalization~\cite{IofSze15} is widely used in deep learning architectures.
Formally, let us consider an input $\Z^l_c$ at layer $l$ and channel~$c$. The output of batch normalization at inference time is a linear function:
\begin{equation}
\label{eq:batch_norm}
    bn(\Z^l_c) = \frac{\Z^l_c - \hat{\mu}^l_c}{\sqrt{\hat{\sigma}^l_c + \epsilon}} \cdot \gamma^l_c + \beta^l_c
\end{equation}  %
Note that all operations are element-wise and $\epsilon$ ensures numerical stability.
The parameters $\gamma^l_c \in \mathbb{R}$ and $\beta^l_c \in \mathbb{R}$ are learnable parameters obtained during training. The parameters $\hat{\mu}^l_c \in \mathbb{R}$ and $\hat{\sigma}^l_c \in \mathbb{R}$ are running estimates of mean and variance over the training batches. They are updated during training and usually fixed at inference time. 
Note that each channel has distinct normalization parameters.

\paragraph{\cnr Sequence}\label{sec:bn-relu}
The Rectified Linear Unit (ReLU) activation function \mbox{$\nonlinear(x)= max(0,x)$} is a standard choice in computer vision models such as ResNet~\cite{HeZhaRen16a} and DenseNet~\cite{HuaLiuVan17}.
It is commonly preceded by batch normalization of convolutional and pooling layer outputs.
As we will discuss in Section~\ref{sec:main}, this \emph{\cnr sequence} $\nonlinear\,(\bn\,(\x))$ allows for simple energy-latency attacks.

\subsection{Efficiency in Machine Learning}
Different techniques have been developed to optimize the training and inference of machine learning models~\cite{NikMahMos18}. Techniques exploiting sparsity are one of the most common directions~\cite{KurKopGel20}. 
Recall that the core computation in neural networks is the multiplication $w \cdot a$ where $w$ is a weight and $a$ an activation. If one of the operands is zero, we can skip the computation and reduce memory requirements. 

As a result, a first direction involves eliminating unimportant weights~$w$, typically done by pruning methods~\cite{HanPooTra15}. %
Fine-grained pruning of specific weights requires some acceleration technique. For example, NVIDIA's A100 GPU can accelerate 2:4 sparsity if every 2 out of 4 weights in a matrix are zero~\cite{MisLatPoo21}. 

In this work, we focus on the second direction, the \emph{activation sparsity}.
This type of sparsity is primarily driven by the choice of activation function. For vision models, ReLU is a standard choice which zeros out negative values. This leads to zero-valued activations, which can be exploited to save energy and to improve latency.  
For example, compression of sparse activation maps allows for more memory-efficient gradient calculation~\cite{JaiPhaMar18}, as well as efficient movement between CPU and GPU using memory virtualization, avoiding performance bottlenecks if data movement is slower than computation~\cite{RhuOCoCha18}.
Moreover, sparsity-aware acceleration strategies~\cite{JiaHuaYu23, JanLeeKim21, OhSoKim21, KurKopGel20, MahEdoZad20, ShaParSud18, ParRhuMuk17} can take advantage of activation sparsity directly on the hardware level to skip ineffectual computations and to reduce memory usage. They often require considering hardware and model architecture together.
Finally, software support for sparsity can provide performance gains. For example, \citet{KurKopGel20} present a convolution algorithm that exploits activation sparsity and that can be used on CPUs.

Unfortunately, activation sparsity is not robust against manipulation. Activations are input-dependant and thus a possible target for attacks at inference time.

\subsection{Energy-Latency Attacks}
\label{sec:sponge_examples}
\begin{figure}
    \centering
    \captionsetup[subfigure]{labelformat=empty}
    \subfloat[][]{
        \centering
        \resizebox{0.45\textwidth}{!}{\includegraphics{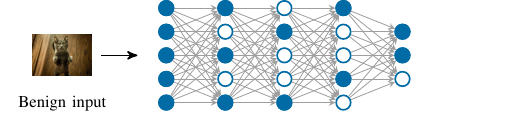}}
    } 
    \\ \vspace{-0.7cm}
    \subfloat[][]{
        \centering
        \resizebox{0.45\textwidth}{!}{\includegraphics{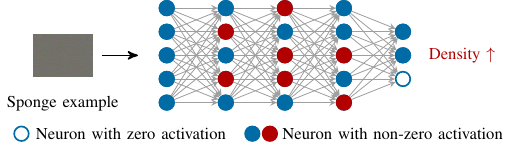}}
    }
    \vspace{-1em}
    \caption[Sponge example]{Illustration of energy-latency attack. Compared to benign inputs, adversarially crafted sponge examples increase the activation density, \ie, the number of non-zero neurons. Thus, sparsity-aware acceleration strategies become ineffective.}
    \label{fig:sponge_example}
\end{figure}

Recent work has demonstrated that attackers can increase the energy and the latency at inference time by submitting specifically crafted input samples---so called \emph{sponge examples}~\cite{ShuZhaBat21}. 
For vision models in particular, the attacker can craft an input that increases the \emph{activation density} by increasing the number of non-zero values across different activation layers. 
\autoref{fig:sponge_example} illustrates this attack principle. 
In this way, attackers can thwart the applicability of acceleration strategies that take advantage of activation sparsity to compress or skip zero values. Gained efficiency savings are therefore reverted. As a result, the energy and latency increases.

Two strategies have been proposed to craft sponge examples~\cite{ShuZhaBat21}. The first, \emph{Sponge-GA}, leverages a genetic algorithm to evolve a pool of randomly initialized images. The number of zero activations during a forward pass is used to obtain fitness scores. Images are mixed with each other and noise is added in each iteration. The second strategy, \emph{Sponge-L-BFGS}, is based on gradient descent and optimizes the following objective:
\begin{equation}
\label{eq:lbfgs-target}
    -\sum_{a^l \in A} || a^l ||_2 \;
\end{equation}
where $A$ is the set of all activation maps~$a^l$ across the model. 
While Sponge-GA is usable for full and zero-knowledge attacks, Sponge-L-BFGS can only be applied with access to the model in a full-knowledge scenario.
In the next section, we examine the attacks' impact on density. These insights allow us to derive simpler attacks with similar attack efficacy in a fraction of the prior computation~time.

\section{Uniformity \& Energy-Latency Attacks}\label{sec:main}
As described before, sponge examples increase the activation density so that acceleration methods cannot exploit sparsity anymore---thus increasing energy and latency. 
While prior work has demonstrated the applicability of these attacks, there is little understanding about the exact attack mechanism. %
In this section, we analyze the impact of energy-latency attacks on neural networks. Our aim is to find out \emph{how} the attacks accomplish their goal to increase the density. 

\begin{figure}[bt]
    \centering
     \includegraphics{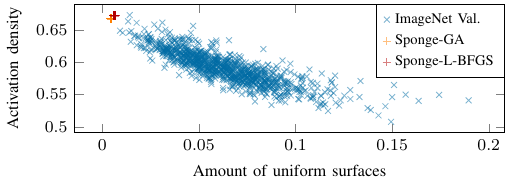}
    \caption{Influence of uniform surfaces on ResNet-18 activation density for 1000 random samples from the ImageNet validation dataset, 100~Sponge-GA samples, and 100~Sponge-L-BFGS samples.}
    \label{fig:uniform_surface_measure}
\end{figure}

\begin{figure}[bt]
    \centering
     \includegraphics{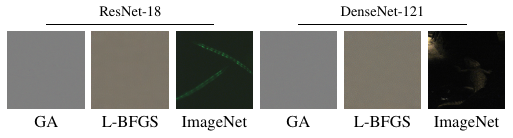}
    \vspace{-0.15em}
    \caption{Examples from Sponge-GA, Sponge-L-BFGS, and ImageNet validation set with highest density on ResNet-18 and DenseNet-121.}
  \label{fig:some_examples_small}
\end{figure}

We start off by noting two observations.
First, activation sparsity almost exclusively occurs after ReLU activation which zeros out negative values. %
Consequently, the attack potential to increase density in turn must lie in these layers.
As batch normalization and convolutional layers are often predecessors, sponge examples need to affect the values in these layers accordingly. 
Second, we observe that sponge examples and the best images from ImageNet with the highest density are all images with flat surfaces of uniform color. \autoref{fig:uniform_surface_measure} depicts the relationship between uniform surfaces and density. The more uniform an input is, the higher the density. We measure the uniformity by computing the standard deviation in a small sliding window that is moved over the image, and finally take the mean over all windows. 
To provide more intuition, \autoref{fig:some_examples_small} shows the best examples from our evaluation.

Consequently, there must be a general interplay between convolution, normalization, and ReLU activation, that makes it beneficial to create uniform sponge examples.


\subsection{Impact of Uniformity on Convolution and Normalization}\label{sec:impact_of_normalization}
Equipped with the prior observations, we can now examine how energy-latency attacks accomplish their goals.
\begin{figure}
  \centering
  \includegraphics[width=\linewidth]{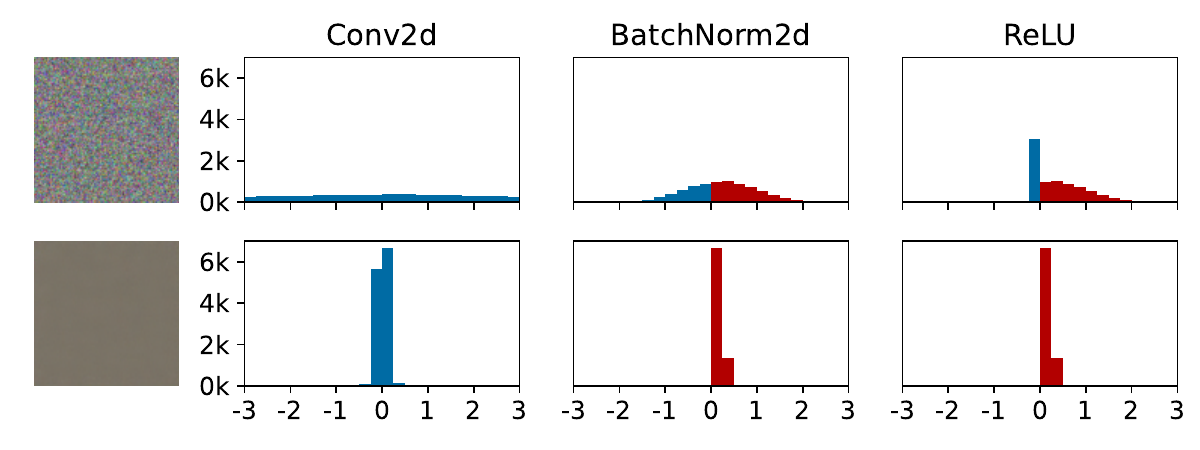}
  \caption{Impact of uniformity. The plot shows the feature-map distributions of a low-density (top) and of a high-density input (bottom) in the first channel of the first sequence of convolution, normalization, and ReLU activation of a ResNet-18 model. The high-density sample is a sponge example obtained with Sponge-L-BFGS. %
  }
  \label{fig:channel_histogram}
\end{figure}
We do this step-by-step using the example in \autoref{fig:channel_histogram}. In our evaluation in \autoref{sec:eval}, we will empirically validate our argumentation. 

\autoref{fig:channel_histogram} shows the respective feature-map distributions after the first sequence of convolution, batch normalization, and ReLU layer for a non-uniform and uniform image, respectively, along the first channel of a ResNet-18 model.
The left column shows that a uniform image leads to a much more concentrated distribution compared to the non-uniform image.
This might not be surprising. If an image has no pattern, a convolution with a kernel---encoding a specific pattern---will trigger a rather small output response. 
Next, batch normalization comes into play. 
The middle column depicts that batch normalization tightens the distribution and slightly shifts it to the right.
As uniform samples cause a tight distribution, even a slight movement to the right can shift the entire distribution to the positive area.
As a consequence, ReLU activation does not set these values to zero (right column in \autoref{fig:channel_histogram}). The activation density increases. In contrast, wider distributions from non-uniform images also shift slightly to the right, but only a fraction of the values move to the positive ReLU region.

Note that a batch normalization can also shift the distribution to the left.
However, we observe that this case occurs less frequently.
To demonstrate this, we calculate the zero thresholds~$\theta^l_c$, indicating when inputs to a batch normalization channel become positive afterwards. %
All parameters are fixed at inference time, so that we can determine general thresholds for each model. 
By setting the function in \autoref{eq:batch_norm} to zero and rearranging the parameters, 
we obtain the zero threshold $\theta^l_c$ as follows:
\begin{align}
    \theta^l_c = \hat{\mu}^l_c - \frac{\beta^l_c \sqrt{\hat{\sigma}^l_c + \epsilon}}{\gamma^l_c} 
\end{align}
If the input $\Z^l_c$ is larger than the zero threshold $\theta^l_c$, it will fall into the positive ReLU region after batch normalization:
\begin{align}
  \bn(\Z^l_c) > 0    \;\;\iff\;\;   \Z^l_c > \theta^l_c 
\end{align}
\begin{figure}
  \centering
  \includegraphics[width=\linewidth]{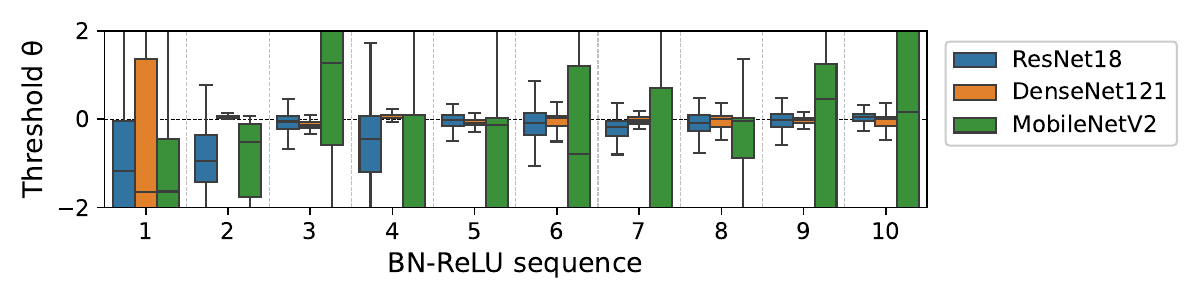}
  \caption[BN Zero Thresholds]{Distribution of zero thresholds~$\theta^l_c$ for the first ten \cnr sequences for different models. Y-axis clipped for better visibility.}
  \label{fig:zero_thresholds}
\end{figure}
We find that the zero threshold is more often negative for common networks such as ResNet. 
\autoref{fig:zero_thresholds} shows the channel-wise distributions of zero thresholds of the first ten \cnr sequences.  
For a tight input distribution around zero, it is then more likely that the whole distribution is shifted to the positive, non-zero area of the ReLU function. The activation density increases.
Since vision models mainly consist of convolutions and pooling layers, tight channel-wise distributions are mostly preserved throughout a forward pass and the described effect takes place at subsequent \cnr sequences.

\subsection{Simple Energy-Latency Attacks} 
\label{sec:our-attacks}
Based on the previous insights, we propose two simple strategies to find inputs with high activation density.

\paragraph{Top Natural Images Strategy}
Natural uniform images can achieve high density.
Hence, a simple yet effective attack is to measure the density of a dataset of natural images and to pick the $n$ best samples. This leads to a set inconspicuous, high-density samples. Note that common data augmentations %
can be applied to increase the diversity of samples if required.  

\paragraph{Uniform Sampling Strategy}
High-density sponge examples from Sponge-GA and Sponge-L-BFGS are rather uniform. We mimic this behaviour with a simple heuristic. We generate sponge examples by drawing image pixels from a narrow Gaussian distribution with standard deviation \mbox{$\sigma=\frac{2}{255} \sim{0}$}. We empirically verify that a small $\sigma$ value leads to the best performance across all models (see Figure~\ref{fig:std_mean_influence}).
The mean parameter~$\mu$ has only a very marginal effect. Its optimal value can be found via grid search on a few test samples. 
Finally note that $\sigma=0$ would only create a single sponge example. Repeated inference of the same input could be easily detected~\cite{CheCarWag20}.

\paragraph{Threat Model}
We design our threat model according to prior work~\cite{ShuZhaBat21}. Our attacks work in a \emph{zero-knowledge} scenario without access to the model parameters, architecture, and training data. Yet, we assume that the attacker is able to query the target remotely to get the inference time or to measure the energy usage (which relates to density). Moreover, we also consider a \emph{blind adversary} without query access. As no measurements are possible, an attacker has to craft suitable sponge examples locally and transfer them to the target.

\section{Evaluation}\label{sec:eval}
We proceed with an empirical examination of our analysis and our proposed attacks. After supporting our analysis empirically (\autoref{sec:eval-impact}), we show that our proposed methods to generate sponge examples achieve a density comparable to prior attacks, but at a fraction of computation time (\autoref{sec:eval-asr}). Finally, we show that our generated sponge examples transfer across vision models (\autoref{sec:eval-transfer}).   

\subsection{Experimental Setup}
\paragraph{Target Models and Dataset}
We follow the setup from \citet{ShuZhaBat21} and perform our evaluation on seven image classification models provided by the TorchVision library, namely ResNet-$\lbrace18, 50, 101\rbrace$~\cite{HeZhaRen16a}, DenseNet-$\lbrace121, 161, 201\rbrace$~\cite{HuaLiuVan17}, and MobileNetV2~\cite{SanHowZhu18}. 
To obtain natural images, %
we use the ImageNet dataset~\cite{RusDenSu+15}. 

\paragraph{Metrics}
We use the \emph{post-ReLU density} as metric for the activation density. This focus is also in line with prior work on activation sparsity~\cite{KurKopGel20}. The sparsity value is also directly related to other metrics such as overall density and energy consumption which were reported by \citet{ShuZhaBat21}. 

\begin{figure}[b]
    \centering
    \input{images/std_mean_influence/std_mean_influence.tex}
    \caption{Parameter search for uniform sampling strategy. Left: Averaged density for 100~inputs sampled from Gaussian noise with fixed $\mu=0.5$ and varying $\sigma$. Right: Activation density for 100~inputs sampled from Gaussian noise with fixed $\sigma=\frac{2}{255}$ and varying $\mu$. 
    }
    \label{fig:std_mean_influence}
\end{figure}
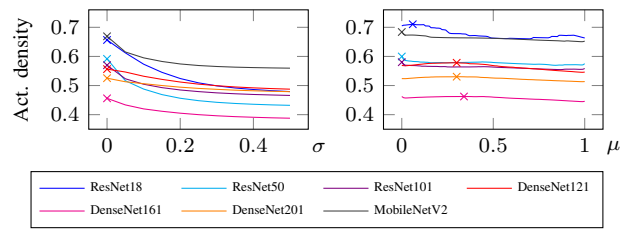

\paragraph{Implementation}
We re-implement Sponge-GA and Sponge-L-BFGS~\cite{ShuZhaBat21}. %
For these attacks, initial inputs are drawn from a uniform distribution $\mathcal{U}(0, 1)$ covering the full pixel range. 
For the Top Natural Images strategy, the ImageNet validation dataset is used to pick the best samples. 
For the Uniform Sampling strategy, we use grid search to identify the optimal parameters %
(see~\autoref{fig:std_mean_influence}).
We set $\sigma=\frac{2}{255}$, and $\mu = 0.1$ for ResNet-18, $\mu = 0.0$ for ResNet-$\lbrace50, 101\rbrace$, DenseNet-161 \& MobileNetV2, and $\mu = 0.3$ for DenseNet-$\lbrace121, 201\rbrace$.
In the following experiments, we report the activation density for 100~sponge examples using Sponge-GA, Sponge-L-BFGS, and our two attack strategies. 
Finally, we include two baselines following prior work~\cite{ShuZhaBat21}. We measure the density on 50,000 images with \emph{random noise}~$\sim~\mathcal{U}(0, 1)$ and on 50,000 images from the ImageNet validation set.

\paragraph{Coherence Check}
We compare the performance of our setup with the previously reported results by \citet{ShuZhaBat21}. 
The difference to our reproduced activation density averaged over all models is $-0.038 \pm 0.015$ and $0.023 \pm 0.050$ for Sponge-GA and Sponge-L-BFGS, respectively. 
We observe only a small difference, indicating a valid experimental setup.

\subsection{Impact Verification}
\label{sec:eval-impact}
We first check our analysis from \autoref{sec:main}. We argued that it is beneficial for sponge examples to be uniform, because this leads to narrow feature map distributions close to zero in the first layers. 
They are more likely to be shifted to the non-zero ReLU region after batch normalization. 

\paragraph{Setup}
We compare high-density and low-density examples. For high density, we consider 100~sponge examples. For low-density, we consider 100~images with lowest density from the ImageNet validation set and 100~random-noise images.  
Due to lack of space, we limit our analysis to ResNet-18, but observe the same effect for other models. 
As the density effect mainly occurs in the first few layers, we further focus on the first ten~\cnr sequences.
Note that all values are recorded at the batch normalization layers. This allows capturing the main execution path of every model and all skip connection channels.
ResNets are a special case. Their skip connections flow from previous ReLU layers and are added to batch normalization outputs. These connections can only add positive or zero values except in rare cases when downsampling occurs. As a result, density can only increase in most \cnr sequences. We still capture the main influence factor for dense activations. 

\begin{figure}
  \centering
  \includegraphics[width=\linewidth]{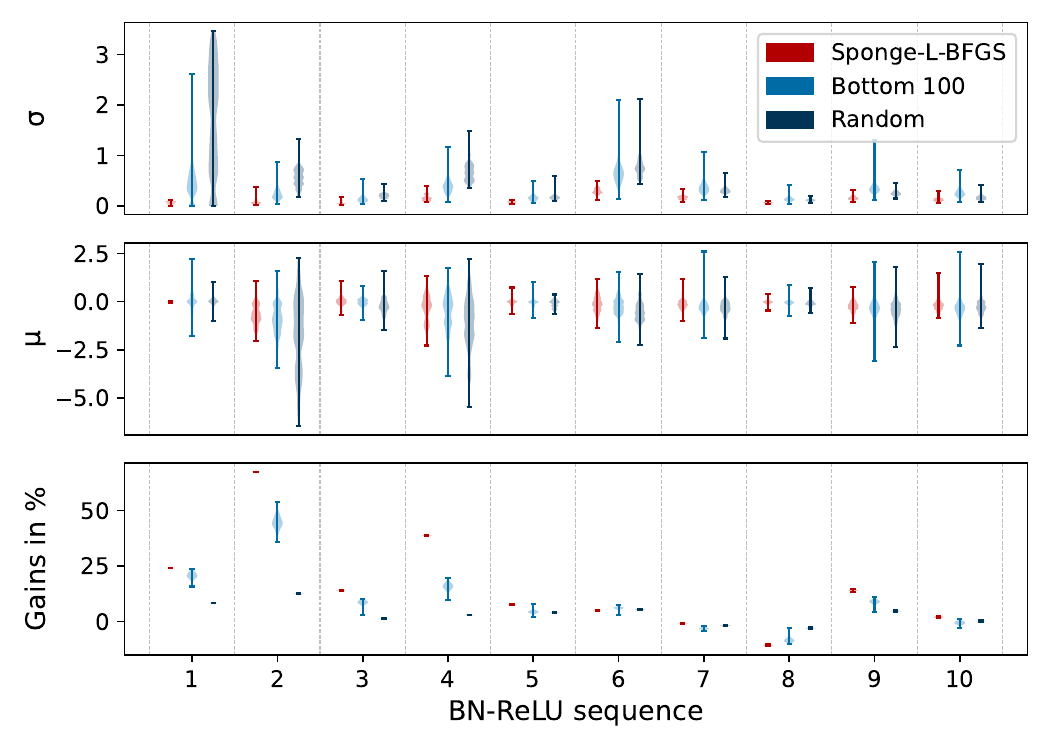}
  \caption[Channel wise Std, Mean, Gain Distributions per Layer]{
  Comparison of higher density (Sponge-L-BFGS) vs. lower density (Bottom-100 and Random) images on ResNet-18. Top and Middle: Standard deviations $\sigma^l_c$ and means $\mu^l_c$ of batch normalization inputs over the channels per \cnr sequence, respectively. Bottom: Distribution of density gains in channels due to batch normalization.
  }
  \label{fig:channel_distributions}
\end{figure}

\paragraph{Results}
\autoref{fig:channel_distributions} captures the results of our analysis from three aspects.
First, the upper plot shows the channel-wise standard deviation of the batch-normalization inputs as violin plot.
Higher-density samples lead to tighter distributions in each channel. 
Second, we also compute the channel-wise mean of the batch-normalization inputs as violin plot (middle plot). The values from high-density samples are more concentrated around zero. 
Third, we consider input \& output channels together and compute a channel-wise \emph{gain} which is the difference of the non-zero ratio after and before applying batch normalization. Sponge examples have higher gains.

Taken together, these three plots confirm our analysis. Sponge examples yield narrower batch-normalization inputs close to zero for each channel. 
From~\autoref{fig:zero_thresholds}, we know that zero thresholds $\theta^l_c$ tend to be negative. Thus, the inputs are likely to be larger, so that the batch-normalization outputs become positive more often. This is reflected by the higher density gains with sponge examples. The ReLU function cannot zero out activations, and the activation density increases.

\subsection{Attack Success Rate}
\label{sec:eval-asr}
We evaluate our proposed energy-latency attacks in terms of their density effect and the required computation time.

\autoref{tab:results_single_column} shows the activation density for the baselines and different sponge-example methods, averaged over 100~images. 
\textit{Our proposed strategies achieve a density effect which is comparable or higher than the baselines and prior work.} 
Note that we omit the standard deviation here, which is close to 0. 

Next, we evaluate the time required to generate sponge examples with each attack strategy. The test system consists of AMD Epyc 7542 CPUs and NVIDEA A30 GPUs. We report the average time for 100 sponge examples. For the Top Natural Images strategy, we measure the time required to forward pass the ImageNet validation set once, and then divide this time by the number of chosen samples ($n=100$ here).

\autoref{tab:times_single_column} shows the average time needed to obtain a sponge example. \textit{Our attacks operate at a fraction of the time that the prior methods Sponge-GA and Sponge-L-BFGS require}.

\begin{table}[t]
\centering
\input{tables/eval_density.tex}

\caption{Average activation density with different strategies (random data and ImageNet validation set baselines, Sponge-GA, Sponge-L-BFGS, Top Natural Images, Uniform Sampling) of different models.}
\label{tab:results_single_column}
\end{table}

\begin{table}[t]
\centering
\input{tables/eval_timing.tex}

\caption{Average time (minutes) required for obtaining a single sample with Sponge-GA, Sponge-L-BFGS, Top Natural Images, Uniform Sampling on different models.}
\label{tab:times_single_column}
\end{table}

\subsection{Transferability}
\label{sec:eval-transfer}
Finally, we analyze if a sponge example---generated on one model---is also effective on another model. This is relevant for blind adversaries who have no query access to the model (recall the threat model in \autoref{sec:our-attacks}). 
To measure the transferability, we compute the percentual increase in activation density of an attack sample crafted for a specific model (y-axis) measured on a target model (x-axis) when compared to a baseline. Baselines are taken by measuring the mean activation density across the ImageNet validation dataset on the target model. 
Results are averaged over 100 attack samples.

\begin{figure}
  \centering
  \includegraphics[width=\linewidth]{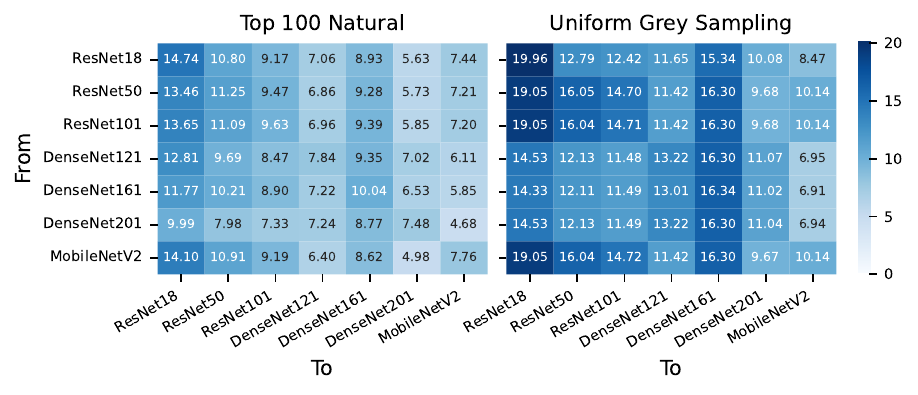}
  \caption[Transferability]{Transferability of our attack strategies across models.}
  \label{fig:transferability}
\end{figure}

\autoref{fig:transferability} shows the transferability property of our proposed strategies. 
\textit{Sponge examples crafted on one model are effective on other models.}
This is expected, since all models have \cnr sequences. 

Taking a closer look, we find that densities for natural images consistently correlate on a \emph{per-sample-basis} across models---in addition to the prior observation of a class-wise transferability~\cite{ShuZhaBat21}.
We measure densities for all samples of the ImageNet validation dataset on all models and calculate Kendall's $\tau$ coefficient~\cite{Ken38} pair-wise across models. The average $\tau$ coefficient (excluding same model pairs) is $0.636$, indicating a similar density per sample across models. 

\section{Discussion}\label{sec:discuss}

\paragraph{Limits of Sponge Examples}
We first note that our analysis on sponge examples does not imply that there is no other method to achieve a comparable or higher density effect. Our work shows that is just beneficial to create uniform samples.
Yet, our analysis pinpoints to two challenges for more effective attacks. First, controlling a distribution in deeper layers based on the input only is difficult. The input impact decreases with deeper layers. Prior work has observed this effect~\cite{ShuZhaBat21} and we can confirm it empirically. 
Second, we also identify a limiting factor in the beginning across the channel dimension. To trigger non-zeros perfectly, an input would have to match all convolutional kernels accordingly. Yet, the kernels react to different patterns, making it difficult to find an adequate input. 
Overall, our analysis underlines that future attacks need to optimize the activations more precisely than just maximizing them. 

\paragraph{Applications For the Good}
We finally discuss how our findings can be leveraged in the opposite direction to improve sparsity and efficiency.
As discussed in \autoref{sec:main}, energy-latency attacks reveal that uniform inputs cause narrow feature-map distributions that cause higher density. Uniform inputs are not necessarily rare, for example, images showing marine life under water or images with the sky in the background.
If a pre-trained model shall only be used for a subset of classes which consist of uniform inputs typically, our results emphasize that some adaption to the subset can provide sparsity improvements. 
We provide preliminary insights here by making a small experiment. We take a pre-trained ResNet 18 model and randomly sample 2,510 training and 216 validation images from ImageNet with uniformity smaller than $0.1$ (see~\autoref{sec:main}), respectively. We fine-tune the model using the uniform training images and test the sparsity on the uniform validation images. The sparsity increases by 4.5\% on average. We also observe that the negative zero thresholds of batch normalization slightly increase. As a sanity check, we randomly sample the same image number from the ImageNet training set and fine-tune ResNet18 with the same setup. This leads to no sparsity gain. While a narrow scenario, our results show that sparsity can be improved for common models with specific data distributions, motivating future work here.

\section{Related Work}\label{sec:relatedwork}
Energy-latency attacks are a new type of attack against learning-based systems, so that there is only little related work. 
\citet{ShuZhaBat21} are the first to show attacks exploiting sparsity-aware systems. We explore these attacks in more detail and derive simpler and faster attack strategies.
\citet{CinDemBig23} study energy-latency attacks on image classifiers at training time. They derive a poisoning attack for an image classifier so that its activation density increases on most inputs. One of their motivation is to avoid the high computation cost by the prior methods Sponge-GA and Sponge-L-BGFS. In this regard, our work shows that much faster attacks are possible. 
In a broader context, energy-latency attacks have also been examined for other domains and applications.
Language systems~\cite{ShuZhaBat21, BouShuAnd22, CheLiuHaq22}, neural image caption generation~\cite{CheSonHaq22}, and object detection in autonomous vehicles~\cite{ShaZolDem23} are vulnerable to energy-latency attacks too. Moreover, attacks have been studied for input-adaptive model architectures~\cite{HanHuaSon22, HaqYan23}, such as multi-exit neural networks~\cite{HaqChaLiu20, HonKayMod21} and neural ODE (Ordinary Differential Equation) models~\cite{HaqCheHaq23}. \citet{CheCheHaq23} demonstrate an attack by injecting a backdoor trigger on input-adaptive models.

\section{Conclusion}\label{sec:conclusion}
Our work examines how sponge examples can reduce activation sparsity. 
We show that uniform inputs are beneficial by causing narrow feature-map distributions that are more prone to be moved to the non-zero (``non-sparse'') ReLU region after batch normalization. The density increases, and so do the energy \& latency. Based on these insights, we then propose two novel attack strategies that are simple, effective, and fast. The sponge examples also transfer across models.

\section*{Acknowledgment}
This work was funded by the Deutsche Forschungsgemeinschaft (DFG, 
German Research Foundation) under Germany's Excellence Strategy -- 
EXC 2092 CASA -- 390781972.
Moreover, this work was supported by fellowships (IFI program and Forschungsstipendien für Doktorandinnen und Doktoranden) of the German Academic Exchange Service (DAAD) funded by the 
Federal Ministry of Education and Research (BMBF).

\interlinepenalty=10000
\bibliographystyle{abbrvnat}
\footnotesize \setlength{\bibsep}{2pt plus 0.5ex}

\normalsize

\end{document}

%% file: commands.tex
\newcommand{\ie}{i.\,e.}

\newcommand{\Z}{\ensuremath{x}\xspace}  %

\newcommand{\nonlinear}{\ensuremath{g}\xspace}  %
\newcommand{\bn}{{bn}\xspace}  %
\newcommand{\x}{\ensuremath{x}\xspace}  %
\newcommand{\ch}{\ensuremath{c}\xspace}  %
\newcommand{\cnr}{BN-ReLU\xspace} %

%% file: tikzpgf.tex
\usepackage{pgfplots} %

\usepackage{pgf}
\usepackage{adjustbox} %

\usepackage{tikz}
\usepackage{pgfgantt}

\pgfplotsset{compat=1.13}
\usetikzlibrary{calc}
\usetikzlibrary{positioning,arrows,fit,matrix}
\usetikzlibrary{backgrounds}
\usetikzlibrary{patterns}
\usetikzlibrary{shapes}
\usetikzlibrary{shapes.arrows} %
\usetikzlibrary{chains}
\usepgfplotslibrary{groupplots}
\usepgfplotslibrary{statistics}
\usepgfplotslibrary{fillbetween}

\tikzset{
	font=\footnotesize,
	connectsimple/.style={thick, shorten >= 0.1em, shorten <= 0.1em},
	connect/.style={->,>=stealth',connectsimple},
}

%% file: images/std_mean_influence/std_mean_influence.tex
\begin{tikzpicture}
    \begin{groupplot}[
            group style={group size= 2 by 1},
            height=3.17cm, width=4.5cm,  %
            ymin=0.35, ymax=0.76,
            x label style={at={(axis description cs:0.975,-0.15)},anchor=west},
        ]
        
        \nextgroupplot[
            ylabel=Act. density,
            xlabel=$\sigma$,
        ]
        \addplot [mark=none, color=blue] table [x=std, y=post_relu_activation_density_ratio, col sep=comma] {images/std_mean_influence/data/std_influence_ResNet18.csv};
        \addplot [mark=x, color=blue] table [x=std, y=post_relu_activation_density_ratio, col sep=comma] {images/std_mean_influence/data/std_influence_max_ResNet18.csv};
        \addplot [mark=none, color=cyan] table [x=std, y=post_relu_activation_density_ratio, col sep=comma] {images/std_mean_influence/data/std_influence_ResNet50.csv};
        \addplot [mark=x, color=cyan] table [x=std, y=post_relu_activation_density_ratio, col sep=comma] {images/std_mean_influence/data/std_influence_max_ResNet50.csv};
        \addplot [mark=none, color=violet] table [x=std, y=post_relu_activation_density_ratio, col sep=comma] {images/std_mean_influence/data/std_influence_ResNet101.csv};
        \addplot [mark=x, color=violet] table [x=std, y=post_relu_activation_density_ratio, col sep=comma] {images/std_mean_influence/data/std_influence_max_ResNet101.csv};
        \addplot [mark=none, color=red] table [x=std, y=post_relu_activation_density_ratio, col sep=comma] {images/std_mean_influence/data/std_influence_DenseNet121.csv};
        \addplot [mark=x, color=red] table [x=std, y=post_relu_activation_density_ratio, col sep=comma] {images/std_mean_influence/data/std_influence_max_DenseNet121.csv};
        \addplot [mark=none, color=magenta] table [x=std, y=post_relu_activation_density_ratio, col sep=comma] {images/std_mean_influence/data/std_influence_DenseNet161.csv};
        \addplot [mark=x, color=magenta] table [x=std, y=post_relu_activation_density_ratio, col sep=comma] {images/std_mean_influence/data/std_influence_max_DenseNet161.csv};
        \addplot [mark=none, color=orange] table [x=std, y=post_relu_activation_density_ratio, col sep=comma] {images/std_mean_influence/data/std_influence_DenseNet201.csv};
        \addplot [mark=x, color=orange] table [x=std, y=post_relu_activation_density_ratio, col sep=comma] {images/std_mean_influence/data/std_influence_max_DenseNet201.csv};
        \addplot [mark=none, color=darkgray] table [x=std, y=post_relu_activation_density_ratio, col sep=comma] {images/std_mean_influence/data/std_influence_MobileNetV2.csv};
        \addplot [mark=x, color=darkgray] table [x=std, y=post_relu_activation_density_ratio, col sep=comma] {images/std_mean_influence/data/std_influence_max_MobileNetV2.csv};

        \nextgroupplot[
            xlabel=$\mu$,
            legend entries={
                ResNet18,
                ResNet50,
                ResNet101,
                DenseNet121,
                DenseNet161,
                DenseNet201,
                MobileNetV2
            },
            legend cell align=left,
            legend columns=4,
            legend style={
                /tikz/every even column/.style={
                    column sep=5pt,
                },
                font=\tiny,
            },
            legend to name={plot:legend},
        ]
            \addplot [mark=none, color=blue] table [x=mean, y=post_relu_activation_density_ratio, col sep=comma] {images/std_mean_influence/data/mean_influence_ResNet18.csv};
            \addplot [mark=none, color=cyan] table [x=mean, y=post_relu_activation_density_ratio, col sep=comma] {images/std_mean_influence/data/mean_influence_ResNet50.csv};
            \addplot [mark=none, color=violet] table [x=mean, y=post_relu_activation_density_ratio, col sep=comma] {images/std_mean_influence/data/mean_influence_ResNet101.csv};
            \addplot [mark=none, color=red] table [x=mean, y=post_relu_activation_density_ratio, col sep=comma] {images/std_mean_influence/data/mean_influence_DenseNet121.csv};
            \addplot [mark=none, color=magenta] table [x=mean, y=post_relu_activation_density_ratio, col sep=comma] {images/std_mean_influence/data/mean_influence_DenseNet161.csv};
            \addplot [mark=none, color=orange] table [x=mean, y=post_relu_activation_density_ratio, col sep=comma] {images/std_mean_influence/data/mean_influence_DenseNet201.csv};
            \addplot [mark=none, color=darkgray] table [x=mean, y=post_relu_activation_density_ratio, col sep=comma] {images/std_mean_influence/data/mean_influence_MobileNetV2.csv};

            \addplot [mark=x, color=blue] table [x=mean, y=post_relu_activation_density_ratio, col sep=comma] {images/std_mean_influence/data/mean_influence_max_ResNet18.csv};
            \addplot [mark=x, color=cyan] table [x=mean, y=post_relu_activation_density_ratio, col sep=comma] {images/std_mean_influence/data/mean_influence_max_ResNet50.csv};
            \addplot [mark=x, color=violet] table [x=mean, y=post_relu_activation_density_ratio, col sep=comma] {images/std_mean_influence/data/mean_influence_max_ResNet101.csv};
            \addplot [mark=x, color=red] table [x=mean, y=post_relu_activation_density_ratio, col sep=comma] {images/std_mean_influence/data/mean_influence_max_DenseNet121.csv};
            \addplot [mark=x, color=magenta] table [x=mean, y=post_relu_activation_density_ratio, col sep=comma] {images/std_mean_influence/data/mean_influence_max_DenseNet161.csv};
            \addplot [mark=x, color=orange] table [x=mean, y=post_relu_activation_density_ratio, col sep=comma] {images/std_mean_influence/data/mean_influence_max_DenseNet201.csv};
            \addplot [mark=x, color=darkgray] table [x=mean, y=post_relu_activation_density_ratio, col sep=comma] {images/std_mean_influence/data/mean_influence_max_MobileNetV2.csv};
    \end{groupplot}

        \node [
            anchor=north,
        ] at ($(group c1r1.outer south west)!0.5!(group c2r1.outer south east)$)
                {\ref{plot:legend}};
\end{tikzpicture}

%% file: tables/eval_density.tex
\begin{tabular}{@{}lllllll@{}}
\toprule
& \multicolumn{2}{c}{Baseline} & \multicolumn{2}{c}{Prior Methods}    & \multicolumn{2}{c}{Our Methods}           \\
\cmidrule(lr){4-5} \cmidrule(lr){6-7}
                & Rand & Val & GA             & L-B            & Top100         & Uni   \\ \midrule
ResNet18        & 0.495 & 0.592 & 0.668 & 0.672 & 0.680 & \textbf{0.710} \\
ResNet50        & 0.440 & 0.517 & 0.593 & 0.571 & 0.575 & \textbf{0.600} \\
ResNet101       & 0.473 & 0.506 & \textbf{0.584} & 0.551 & 0.555 & 0.580 \\
DenseNet121     & 0.497 & 0.511 & \textbf{0.581} & 0.576 & 0.551 & 0.578 \\
DenseNet161     & 0.395 & 0.397 & \textbf{0.467} & 0.464 & 0.437 & 0.462 \\
DenseNet201     & 0.486 & 0.478 & 0.551 & \textbf{0.556} & 0.513 & 0.530 \\
MobileNetV2     & 0.564 & 0.621 & 0.672 & 0.651 & 0.669 & \textbf{0.684} \\
\end{tabular}

%% file: tables/eval_timing.tex
\begin{tabular}{@{}lllllll@{}}
\toprule
& \multicolumn{2}{c}{Prior Methods} & \multicolumn{2}{c}{Our Methods}           \\
\cmidrule(lr){2-3} \cmidrule(lr){4-5}
& GA     & L-BFGS         & Top100  & Uni  \\ \midrule
ResNet18        & 24.59  & 27.50  & 0.01 & \textbf{0.00}   \\
ResNet50        & 51.24  & 26.58  & 0.04 & \textbf{0.00}   \\
ResNet101       & 71.11  & 61.52  & 0.06 & \textbf{0.00}   \\
DenseNet121     & 82.71  & 67.16  & 0.06 & \textbf{0.00}   \\
DenseNet161     & 105.56 & 103.03 & 0.13 & \textbf{0.00}   \\
DenseNet201     & 100.46 & 136.49 & 0.13 & \textbf{0.00}   \\
MobileNetV2     & 50.02  & 22.77  & 0.03 & \textbf{0.00}   \\
\end{tabular}